\documentstyle[prl,aps,graphicx,multicol]{revtex}

\begin{document}
\draft
\title{The Nature of Heavy Quasiparticles in Magnetically Ordered Heavy Fermions}
\author{M. Dressel$^{1}$\cite{email}, N. Kasper$^{1}$, K. Petukhov$^{1}$, B. Gorshunov$^{1}$,
 G. Gr{\"u}ner$^{2}$,
 M. Huth$^{3}$, and H. Adrian$^{3}$}
\address{$^{1}$1.~Physikalisches Institut, Universit\"at Stuttgart, Pfaffenwaldring 57, D-70550 Stuttgart, Germany\\
$^{2}$Department of Physics, University of California, Los Angeles,  CA, 90095-1547\\
$^{3}$Institut f\"ur Physik, Universit\"at Mainz, D-55099 Mainz, Germany}
\date{Received  \today}
\maketitle

\begin{abstract}
The optical conductivity of 
the heavy fermions UPd$_2$Al$_3$ and UPt$_3$ has been 
measured in the frequency range from 
10 GHz to 1.2 THz (0.04 meV to 5 meV) 
at temperatures $1~{\rm K}<T< 300$~K. 
In both compounds  a well pronounced pseudogap
of less than a meV develops in the optical response at low temperatures; 
we relate this to the antiferromagnetic ordering.
From the energy dependence of the effective electronic mass and 
scattering rate we derive the energies essential for the heavy quasiparticle.
We find that the enhancement of the mass mainly occurs 
below the energy  which is related to magnetic correlations between
the local magnetic moments and the itinerant electrons. 
This implies that the magnetic order in these compounds is
the pre-requisite to the formation of the  heavy quasiparticle 
and eventually of superconductivity.
\end{abstract}


\begin{multicols}{2}
\columnseprule 0pt
\narrowtext

In common metals itinerant electrons of the conduction band
are responsible for the electrical transport.
The conductivity is frequency independent until the scattering with phonons and imperfections leads
to a drop in the real part of the optical conductivity $\sigma_1(\omega)$ 
around the scattering frequency $\Gamma$.
This behaviour is well described by the Drude model:
$\hat{\sigma}(\omega)=\sigma_1+i\sigma_2={(Ne^2/m)}/({\Gamma-i\omega})$ with the 
charge carrier mass $m$ and scatting rate $\Gamma$  assumed to be frequency
independent. Here $N$ is the carrier concentration,
 $e$ is the electronic charge, and $\omega$ is the frequency \cite{DresselGruner}.
While at room temperature also heavy-fermions metals follow this scenario, 
at liquid-helium temperatures large deviations are observed which are
caused by many-body effects. 

Heavy fermions show an unusual interplay of electronic
correlations due to an interaction of itinerant electrons and 
local magnetic moments. At low temperatures these intermetallic compounds exhibit 
significant increase of the magnetic susceptibility and the electronic contribution
to the specific heat compared to most metals. 
This is commonly explained by an enhanced effective mass  $m^*$ 
of some hundred times the free electron mass due to
the highly correlated electronic behaviour and is described in the context of interacting
Fermi liquids \cite{Fisk88}. 
The strong interaction of the quasi-free
conduction-band electrons with nearly localized $f$-electrons leads to a many-body resonance, i.e., an enhanced 
density of states at the Fermi energy \cite{Grewe91}. 
The pile-up of this
narrow resonance sets in below the so-called coherence temperature $T^*$, which is experimentally determined by
a drop in the dc-resistivity and a cusp in the susceptibility, and
typically is between 10 and 100~K. 

Also the optical properties show clear signatures of the coherent many-body ground state: 
As a consequence of the large effective mass $m^*$ the spectral weight 
$\int\sigma_1(\omega){\rm d}\omega=\omega_p^{*2}/8$ is reduced. 
$\omega^*_p=(\frac{4\pi Ne^2}{m^*})^{1/2}$  is called
the renormalized plasma frequency. The relevant electronic excitations shift
to very low energies, best characterized by an effective scattering rate 
$\Gamma^*$ \cite{Degiorgi99}. 
With decreasing temperature the
electronic correlations become more pronounced, leading to a
gradual increase of the effective mass $m^*$ and the corresponding decrease of the effective scattering rate.
The  interaction of the conduction electron spins with the atomic moments in the heavy-fermion crystals explains 
the reduction of the magnetic susceptibility and 
also leads to a hybridization gap $\Delta$ in the density of states. 
The gap scales with the coherence temperature $(\Delta/T^*)^2=m/m^*$ and is commonly observed  
in the far infrared range of frequency \cite{Hewson97,Dordevic01}.
Using a frequency dependent scattering rate and effective mass for
the description of  the optical properties of correlated electron systems 
one obtains the energy dependence of the renormalization effects
which goes beyond the information derived from static thermodynamic measurements. In the low-energy limit
the frequency dependence of both parameters $m^*(\omega)$ and $\Gamma(\omega)$ should resemble 
their temperature dependence \cite{DresselGruner}.

An additional aspect becomes important for those heavy-fermion compounds 
which undergo a magnetic phase transition. 
Antiferromagnetism is often due to the ordering of localized magnetic moments;
this commensurate magnetic structure does not lead to a gap at the Fermi energy.
In the case of a spin density wave, 
a spatial density modulation of the spin direction of the conduction electrons leads to antiferromagnetism, 
which in general is incommensurate with the underlying lattice; 
a gap in the electronic density of states at the Fermi energy is observed.
Optical experiments exhibit a clear gap feature in the latter case while 
 for the first case we expect no influence on the optical properties \cite{Degiorgi97}.

Because of its two-component electronic character,
UPd$_2$Al$_3$ exhibits both pronounced local magnetic moment and heavy-fermion itinerant behaviour \cite{Geibel91,Caspary93,Feyerherm94}. 
The $5f$-shell has an average occupation of slightly less than three, with two electrons localized  
in the U$^{4+}$ state and the remaining $5f$-electrons considered to be itinerant due to their large hybridization 
with the conduction electrons \cite{Knopfle96}. 
This coexistence is seen in the large magnetic moment (0.85~$\mu_B$) well below the antiferromagnetic 
ordering temperature $T_N\approx 14$~K 
which is almost atomic-like, 
while at the same time the itinerant electrons display a large Sommerfeld coefficient of the electronic specific heat 
$\gamma=140$~mJ\,mol$^{-1}$K$^{-2}$ which indicates an effective mass $m^*/m\approx 50$. 
The magnetic ordering is of local origin and the formation of a spin density wave 
can be ruled out \cite{Degiorgi97,Krimmel96}.
The heavy quasiparticles condense in the superconducting state below $T_c\approx 2$~K.
In order to reconcile magnetic ordering and superconductivity
it was first suggested that different parts
of the Fermi surface lead to antiferromagnetism and to the heavy-fermion ground  state. 
Recent observations by tunneling spectroscopy and inelastic neutron scattering indicate, 
however, that magnetic excitons 
produce the effective coupling between the itinerant electrons and 
are responsible for superconductivity in UPd$_2$Al$_3$ \cite{Jourdan99,Sato01}. 
 
If there is  an interaction between
magnetic and electronic excitations in the superconducting state, 
the question arises, whether the magnetic ordering also affects the heavy quasiparticles in the normal state.
In which way, or how far, is the low-temperature ground state (still $T>T_c$), 
where the large effective mass $m^*$ has
been determined, affected by the magnetic ordering? 
Are solely the electronic correlations of the $f$-shell electrons with the conduction-band 
electrons responsible for this enhancement as was argued above within the common heavy-fermion picture? 

In order to address these questions, we have investigated the optical properties of 
UPd$_2$Al$_3$ in the frequency range from 10~GHz to 1.2~THz (0.3 to 40~cm$^{-1}$, 
corresponding to a photon energy of 0.04 to 5~meV) at 
temperatures $2~{\rm K}<T<300$~K. 
Using a Mach-Zehnder interferometer equipped with tunable, coherent radiation sources we measured the absorption and 
the phase shift of the radiation
upon passing through a 150~nm thick epitaxial grown
 UPd$_2$Al$_3$-film deposited on a LaAlO$_3$ substrate by molecular beam epitaxy \cite{Huth93}. 
In addition we performed microwave and millimeter wave experiments utilizing enclosed cavities \cite{Dressel02}. 
Our data were supplemented with reflectivity measurements on bulk samples \cite{Degiorgi97} in order to 
obtain the absorptivity over an extremely broad frequency range as plotted in Fig.~1a. 
The lines in this figure represent the input we used for the Kramers-Kronig analysis in order to obtain the 
optical conductivity. 
The result of it is plotted in the lower panel of  Fig.~1;
in addition we display the directly measured conductivity data.
In the Kramers-Kronig analysis tried to fit both the absorptivity and the conductivity data best, 
taking the  different experimental uncertainties into account \cite{Dressel02}. 

The main findings can be summarized as follows:
At high temperatures we observe a broad conductivity spectrum as expected for a Drude metal.
As the temperature decreases below $T^*\approx 50$~K, a renormalized Drude peak develops 
due to the gradual enhancement of the effective mass $m^*$. 
At the same time, a gap feature develops around 100~cm$^{-1}$
as expected from the hybridization of the localized $5f$-electrons
and the conduction electrons. This behaviour
is typical for heavy fermions \cite{Degiorgi99}.
Lowering the temperature  further, magnetic ordering sets in at $T_N$
and leads to distinct signatures in the optical spectrum.
 In the magnetically ordered state  we observe a well pronounced pseudogap below 2~cm$^{-1}$ 
which we assign to the interaction of the localized magnetic moments and spins of the itinerant electrons. 
At even lower frequencies, an extremely narrow zero-frequency mode re\-mains which is responsible for 
the dc conductivity and for superconductivity below $T_c$.

To describe the low-energy data satisfactorily,
we introduce a complex frequency dependent scattering
rate $\hat{\Gamma}(\omega) =
\Gamma_1(\omega) + i\Gamma_2(\omega)$ into the
standard Drude form of $\hat{\sigma}(\omega)$. With  the
dimensionless quantity
$\lambda(\omega) =-\Gamma_2(\omega)/\omega$, the complex
conductivity can be written as~\cite{DresselGruner}
\begin{equation}
\hat{\sigma}(\omega)=\sigma_1 + i\sigma_2= 
\frac{(\omega_p^{\prime})^2}{4\pi}
\frac{1}{\Gamma_1(\omega)-i\omega(m^*(\omega)/m )}
\label{eq:gen-drude2}
\end{equation}
where $m^*/m =1+\lambda(\omega)$ is the frequency dependent enhanced mass. $(\omega_p^{\prime})^2$ 
corresponds to the fraction of electrons
which participate in the many-body state developing below $T^*$.
By rearranging Eq.~(\ref{eq:gen-drude2}) we can write expressions for
$\Gamma_1(\omega)$ and $m^*(\omega)$ in terms of $\sigma_1(\omega)$ and
$\sigma_2(\omega)$ as follows:
\begin{equation}
\Gamma_1(\omega)=\frac{(\omega_p^{\prime})^2}{4\pi}
\frac{\sigma_1(\omega)}{|\hat{\sigma}(\omega)|^2}
\label{eq:gam-w} ,
\end{equation}
\begin{equation}
\frac{m^*(\omega)}{m }=\frac{(\omega_p^{\prime})^2}{4\pi}
\frac{\sigma_2(\omega)/\omega}{|\hat{\sigma}(\omega)|^2}.
\label{eq:mstar-w}
\end{equation}
Such analysis  allows us to look for
interactions which would lead to energy dependent renormalization as
the frequency dependent scattering rate and effective mass are related to 
the real and imaginary parts of the energy dependent self-energy of the electrons \cite{Abrikosov65}.
In Fig.~2 the  frequency dependences of the scattering rate $\Gamma_1(\omega)$ and the effective 
mass $m^*(\omega)$ are displayed for different temperatures.  As expected for a normal metal, at $T>T^*$ 
the spectra of $\Gamma_1(\omega)$ and $m^*(\omega)$ are nearly frequency independent. 
As the temperature is lowered we observe a peak in the energy dependent scattering rate at the 
hybridization gap, as can be nicely seen in the $T=30$~K data of Fig.~2b. The decrease of 
$\Gamma_1(\omega)$ to lower frequencies corresponds to an increase of $m^*(\omega)$. 
At these intermediate temperatures, $T_N<T<T^*$, the effective mass $m^*/m$ gradually 
increases when going to smaller frequencies or correspondingly to lower temperatures; 
eventually it reaches a value of 35. 
Decreasing the temperature below $T_N$ results in a second
peak of $\Gamma_1(\omega)$ and a strong increase of the effective mass below 1~cm$^{-1}$. 
As already seen at $T=15$~K, this effect becomes stronger as the temperature is lowered to 2~K. 
The effective mass $m^*(\omega)$ levels off at $m^*/m\approx 50$ below 1~cm$^{-1}$and 
nicely matches the values obtained by thermodynamic methods \cite{Geibel91} 
which are indicated by the star in Fig.~2c. 
We see a moderate enhancement of $m^*(\omega,T\rightarrow 0)$ for frequency below the 
hybridization gap
and a larger one below the gap which develops 
due to magnetic correlations. Accordingly $m^*(\omega\rightarrow 0,T)$ increases
only slightly below $T^*$ and shows a large enhancement below $T_N$. 
The scattering rate drops more than one order of magnitude upon passing $T_N$;
this can be explained by the freezing-out of spin-flip scattering in the ordered state.
The energy-dependent scattering rate is related 
to the electronic density of states. 
In the case of UPd$_2$Al$_3$ we can distinguish two gap structures with an enhanced density of states at the edges, 
features known for superconductors. 

The estimation of the reduced 
spectral weight $\omega^*_p=c/\lambda_L(0)\approx4000~{\rm cm}^{-1}$ from the London penetration 
depth $\lambda_L =450$~nm \cite{Caspary93} agrees with our results of $4300~{\rm cm}^{-1}$ if we
integrate the conductivity spectrum up to $100~{\rm cm}^{-1}$;
a renormalized Drude behaviour without a gap feature would yield a value more than twice as large. 
This implies that  all carriers seen in our low-energy
spectra are in the heavy-fermion ground state and eventually undergo the
superconducting transition below $T_c$. 
We can definitely rule
out an assignment of the gap simply
to the localized carriers of the antiferromagnetic 
ordered states, with the delocalized carriers contributing only to
the narrow feature at $\omega=0$, because its small
spectral weight at low temperatures accounts for only 18 \%\
of the carriers which become superconducting. This also means that
the excitations above the gap stem from the delocalized states
and that the gap observed in our conductivity spectra is related
to exchange correlations of the second subsystem.
Recently it was found that the magnetic excitations
seen by neutron scattering produce effective interactions between itinerant
electrons and lead to superconductivity \cite{Sato01}. 
 Our results now imply that the magnetic excitations responsible for superconductivity also cause the heavy quasiparticles.

The picture we developed in order to explain the optical properties of UPd$_2$Al$_3$ 
should also apply to other heavy fermion compounds, in particular to UPt$_3$ which exhibits a similar 
behaviour in many regards \cite{Fisk88,Ott87}.  
For UPt$_3$ the effective mass of the quasiparticles 
is larger ($m^*/m\approx 200$) and the relevant energy scales are lower. 
The coherence temperature is $T^*\approx 30$~K, fluctuating short range magnetic order  ($0.02\mu_B$) 
occurs at $T_N=5$~K, and superconductivity sets in at $T_c=0.5$~K. 
Recent bandstructure calculations \cite{Zwicknagl01} infer the existence of localized 
as well as delocalized $5f$-electrons in UPt$_3$ very much similar to UPd$_2$Al$_3$. 
It was suggested that the observed enhancement of the quasiparticle mass results from 
the local exchange interaction of two localized $5f$-electrons with the remaining delocalized ones \cite{Zwicknagl01}. 

We have reanalyzed the optical properties of UPt$_3$ previously obtained by various techniques
 \cite{Sulewski88,Donovan97} in the same way as described above and can identify similar features (Fig.~3). 
When the coherent ground state builds up ($T<T^*$), the optical conductivity 
increases for frequencies below 30~cm$^{-1}$. 
As the temperature drops below $T_N=5$~K, magnetic ordering occurs  
and an energy gap progressively develops at about $3~{\rm cm}^{-1}$ which
is assigned to magnetic correlations \cite{Donovan97}. 
The frequency dependence of the effective mass displayed in Fig.~3c clearly shows 
at the lowest temperature accessible that only a marginal increase of $m^*$ is observed around 
30~cm$^{-1}$, while below the energies related to the magnetic correlations the mass is drastically enhanced.
Thus we also find that, as for UPd$_2$Al$_3$, the coupling of the localized and 
delocalized $5f$-electrons causes the heavy quasiparticles in  UPt$_3$; these 
magnetic excitations are very likely to be responsible for superconductivity.

For both compounds, it therefore appears that the formation of the heavy quasiparticles relies on the
establishment of antiferromagnetic order rather than competition of the coherent singlet formation and
the magnetic order. Since we made similar observations for two materials 
which may have a different ground state as evidenced by the vastly different $\mu_B$,
it remains to be seen how far our findings have implications on non-magnetic heavy 
fermions with a large effective mass.
Thermodynamic measurements  should test our idea
that the effective mass $m^*(T)$ increases only slightly below $T^*$ but 
shows a large enhancement below $T_N$. It also remains to be seen in which 
way our observation is related to the pseudogap 
which was established in high-temperature superconductors well above $T_c$
by optical and other methods \cite{Timusk99}. 

In conclusion, the electrodynamic response of two
uranium-based heavy fermion superconductors with magnetic ordering
exhibits a behavior at low temperatures and
low frequencies which cannot be explained within the simple picture of a
renormalized Fermi liquid.  
The low-energy response shows the well-known
renormalized behavior only above the magnetic ordering temperature; below $T_N$  additional features were discovered.
Besides an extremely narrow  zero-frequency response, we observe a pseudogap of less than a meV. The 
experiments yield indications that this pseudogap is connected to magnetic correlations on the delocalized charge carriers. 
We have argued, that this gap at extremely low energies is due to the 
influence of the localized magnetic moments on the conduction electrons 
and this interaction is mainly responsible for the enhanced mass $m^*$.

We acknowledge helpful discussions with P. Fulde, A. Muramatsu, A. Schwartz, F. Steglich, and G. Zwicknagl. 
The work was supported by the DFG.

\begin{figure}
\caption{\label{fig1}(a) Frequency dependent absorptivity of UPd$_2$Al$_3$ at different temperatures  measured
over a wide frequency range.
The symbols on the left axis represent the
dc values in a Hagen-Rubens behaviour;
the full symbols in the microwave range are obtained by cavity perturbation technique; the open sympols present absorption evaluated from the
transmission and phase measurement by the Mach-Zehnder interferometer 
(only 2~K data are plotted for clarity reasons).
The lines are obtained by combining the various optical investigations (transmission through films and reflection of bulk samples).
(b) Corresponding optical conductivity of UPd$_2$Al$_3$ as evaluated by a Kramers-Kronig analysis of the above absorption data. The data points on the left axis indicate the dc conductivity;
the full symbols correspond to the direct determination of the optical conductivity using the transmission and the phase shift obtained by the Mach-Zehnder interferometer.}
\end{figure}

\begin{figure}
\caption{\label{fig2}Frequency dependence of (a) the optical conductivity $\sigma_1(\omega)$, (b)
the scattering rate 
$\Gamma_1(\omega)$, and (c) the effective mass $m^*(\omega)$ of UPd$_2$Al$_3$ for different temperatures. The 
point on the left axis correpsonds to the effective mass derived by thermodynamic measurements \protect\cite{Geibel91}. The arrow marks the correlation gap.}
\end{figure}

\begin{figure}
\caption{\label{fig3}Frequency dependence of (a) the optical conductivity $\sigma_1(\omega)$, (b) 
the scattering rate 
$\Gamma_1(\omega)$, and (c) the effective mass $m^*(\omega)$ of UPt$_3$ for different temperatures using
the optical data obtained by \protect\cite{Sulewski88,Donovan97}.}
\end{figure}

\end{multicols}
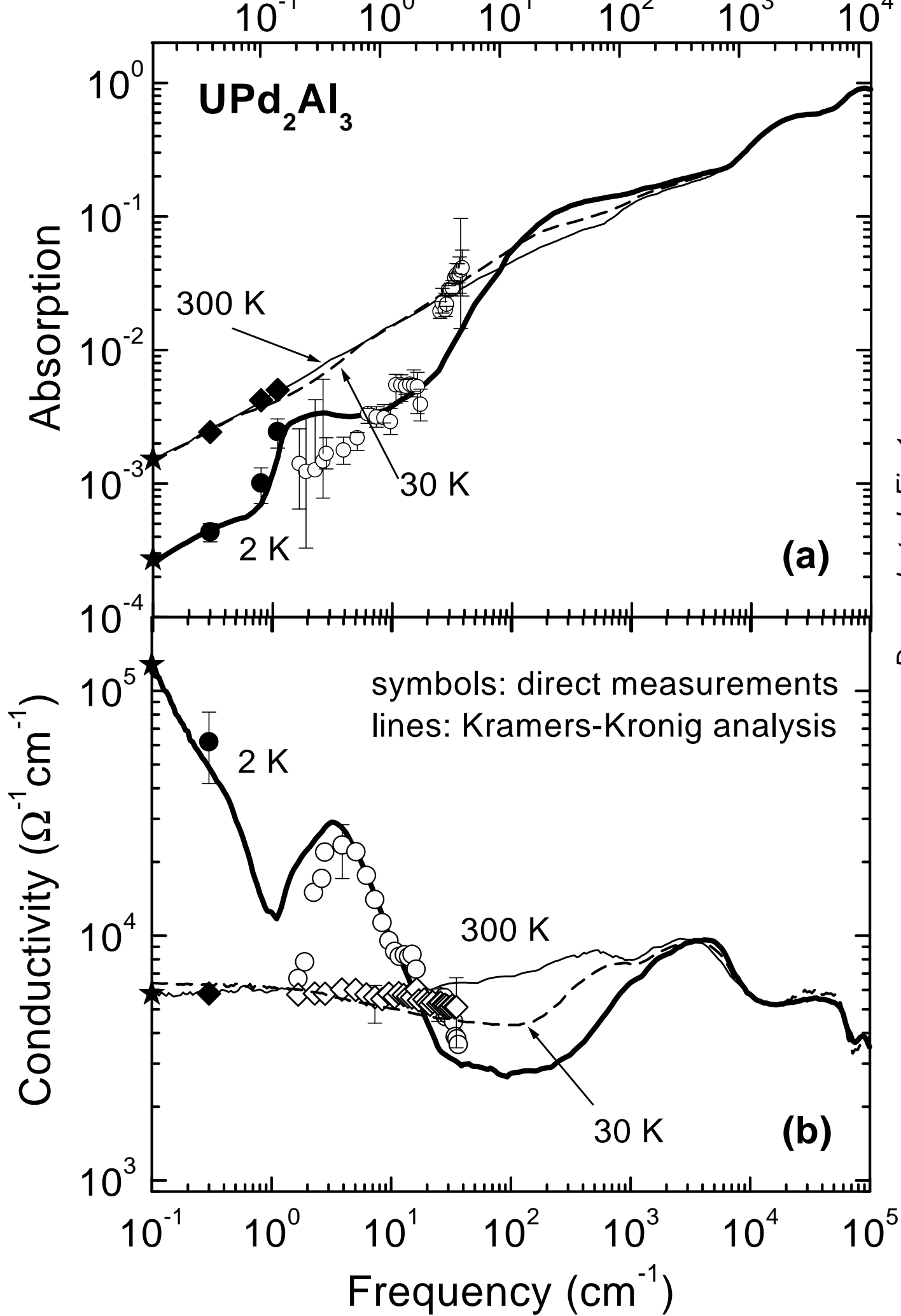
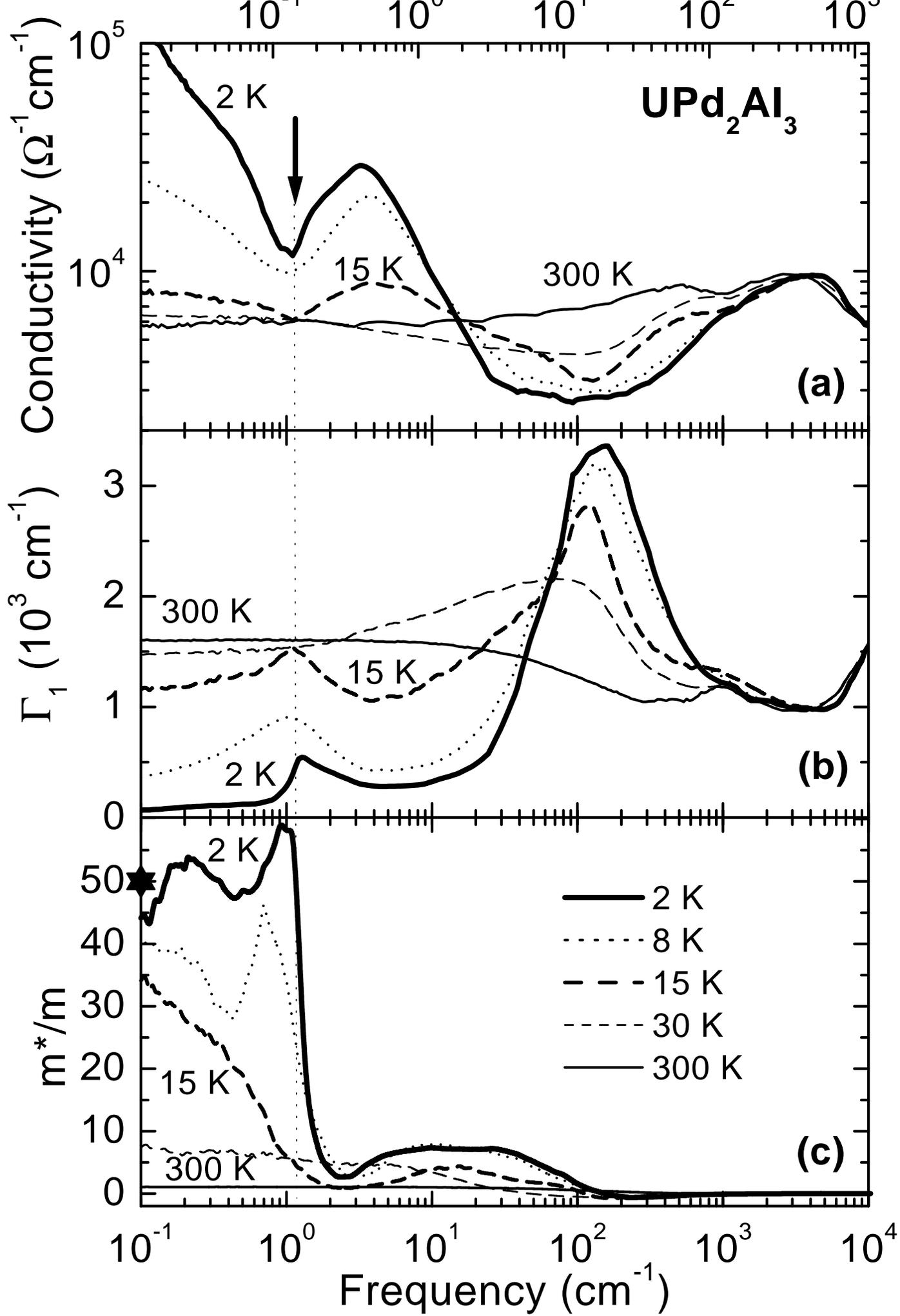
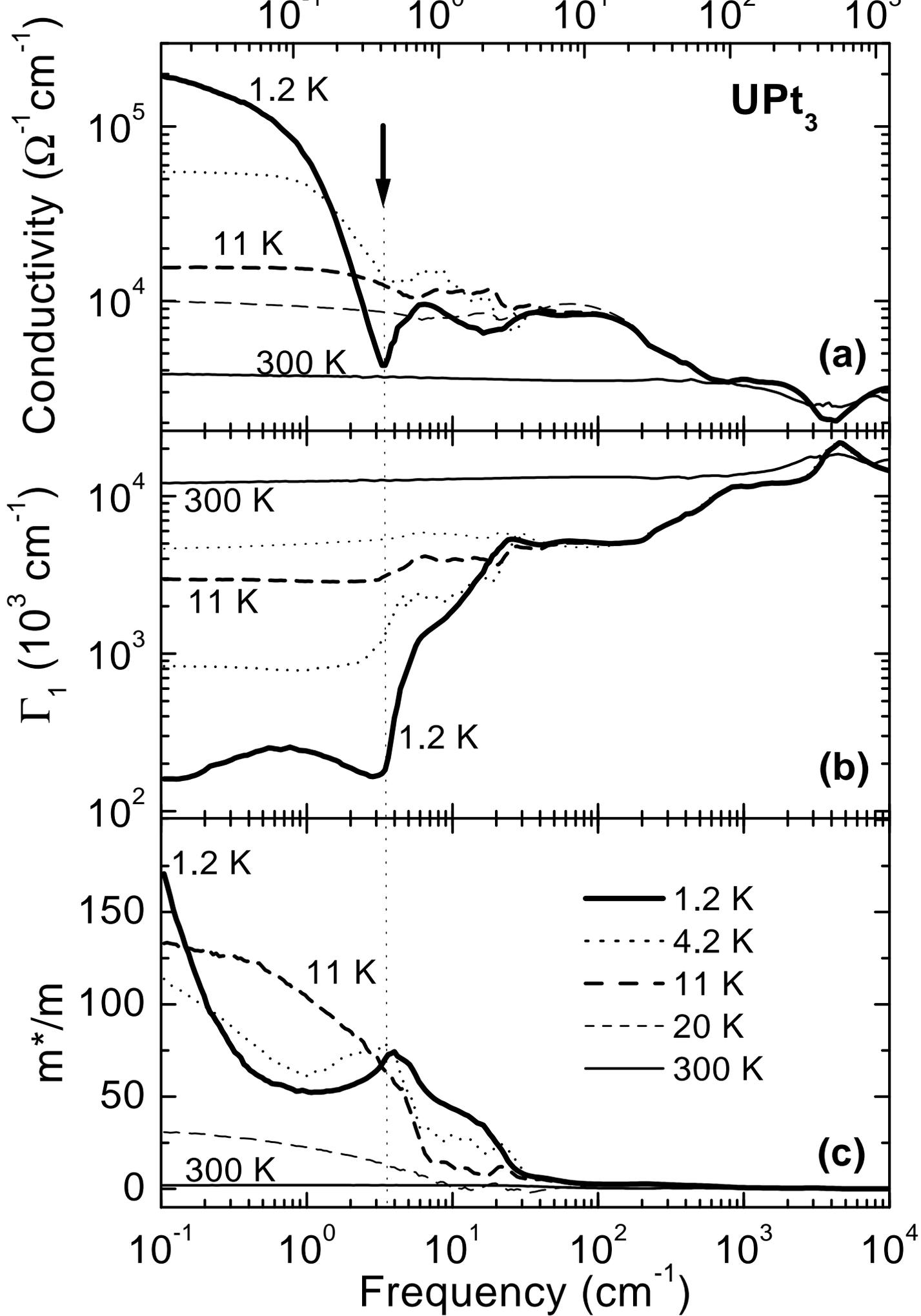
\end{document}